\documentstyle[amssymb,aps]{revtex}

\begin{document}
\title{Castaing Instability and Precessing Domains in Confined Alkali Gases}
\author{A. Kuklov$^{a)}$ and A. E. Meyerovich$^{b)}$}
\address{$^{a)}$Department of Physics, CUNY - Staten Island, New York, NY \\
10314\\
$^{b)}$Department of Physics, University of Rhode Island,\\
Kingston, RI 02881}
\maketitle

\begin{abstract}
We explore analogy between two-component quantum alkali gases and
spin-polarized helium systems. Recent experiments in trapped gases are put
into the frame of the existing theory for Castaing instability in transverse
channel and formation of homogeneous precessing domains in spin-polarized
systems. Analogous effects have already been observed in spin-polarized $%
^{3}He$ and $^{3}He-\,^{4}He$ mixtures systems. The threshold effect of the
confining potential on the instability is analyzed. New experimental
possibilities for observation of transverse instability in a trap are
discussed.
\end{abstract}

\section{Introduction}

Remarkable observation of spontaneous spatial separation of states in recent
experiments on trapped $^{87}Rb$ \cite{jila1} was immediately followed by
three, almost simultaneous theoretical explanations \cite{lev1,lal1,wil1}.
Though these explanations differ in details, all three exploit the analogy
between a two-component $^{87}Rb$\ gas and a spin-polarized spin-1/2 gas.
The analogy is based on the similarity of strong molecular fields in all
two-component quantum gases in the Boltzmann temperature range. In
spin-polarized gases, this molecular field can cause transverse spin
currents \cite{mb1} which, in turn, can result in spatial separation of
spin-up and spin-down components. An analog of this effect was, according to
Refs. \cite{lev1,lal1,wil1}, observed in experiment \cite{jila1} with $%
^{87}Rb$.

The aim of this paper is to point out that the analogy between
multi-component and spin-polarized gases is even richer. In mid-90s there
were several interesting experimental and theoretical publications on
spin-polarized systems that are similar in spirit to Refs. \cite
{jila1,lev1,lal1,wil1}. The key is the so-called Castaing instability of
systems with inhomogeneous polarization with respect to transverse
fluctuations and the formation of precessing magnetic domains. These effects
have been studied, both experimentally and theoretically, in the context of
spin-polarized helium systems. We will analyze the analogs of these effects
for alkali gases and make suggestions for new types of experiments including
the observation of the instabilities. So far, there have been no such
instability experiments for alkali gases. Though the two-component alkali
gases are to a large extent similar to spin-polarized helium, the
characteristic parameters are different and the issue should be revisited in
application, for example, to $^{87}Rb$. The main difference is the small
size and non-uniformity of confined clouds of alkali gases.

\section{Main equations}

The analogy between spin-polarized quantum gases and multi-component gases
(gases of particles with discrete internal degrees of freedom) has been
discussed in detail in Ref. \cite{mey1}. For example, the densities of
components of a two-component gas ({\it e.g., }$^{87}Rb$) play the role of
the spin-up and spin-down components of the spin-polarized gas, while the
(Rabi) transitions between the states are similar to the transitions between
the spin components and lead to the formation of the transverse (to the
quantization axis) magnetization and correspond to the off-diagonal
components of the spin density matrix.

First, let us translate the equations of motion of Ref. \cite{mey1} onto the
language of trapped two-component quantum Boltzmann gas in the $s$-wave
approximation (in experimentally interesting case, the gas temperature is so
low that the particle wavelength is much larger than the scattering length $%
a $). For a two-component gas, the density $\widehat{n}$ is a $2\times 2$
matrix. The diagonal components $n_{1,2}$ of this matrix represent densities
of pure components and obey the conservation law

\begin{equation}
\partial _{t}n_{1,2}({\bf r},t)+{\bf \nabla }_{i}J_{i}^{\left( 1,2\right) }(%
{\bf r},t)=0,  \label{eq10}
\end{equation}
where $J_{i}^{\left( 1,2\right) }$ is the corresponding current. The
off-diagonal (mixed) matrix element $n_{12}$\ describes transitions between
the components and obeys the equation

\begin{equation}
\partial _{t}n_{12}({\bf r},t)-i\delta Un_{12}({\bf r},t)+\nabla
_{i}J_{i}^{\left( 12\right) }({\bf r},t)=I_{12}^{coll},  \label{eq111}
\end{equation}
where 
\begin{equation}
\delta U=U_{1}({\bf r})-U_{2}({\bf r})+\left( g_{11}-g_{22}\right) n({\bf r}%
),  \label{eq12}
\end{equation}
$U_{1,2}$ is the trapping potential for the component $\left( 1,2\right) $,
and the interaction constants $g_{ik}$ in $^{87}Rb$ are expressed via the
corresponding scattering lengths as 
\begin{equation}
g_{ik}=4\pi a_{ik}/m,\ a_{11}=a+a^{\prime },\ a_{22}=a-a^{\prime },\
a_{12}=a.  \label{eq13}
\end{equation}
Since the asymmetry of scattering is weak, $a^{\prime }\ll a,$ the collision
integral $I_{12}^{coll}$, which is proportional to $\left( a^{\prime
}/a\right) ^{2}$, is negligible, $I_{12}^{coll}\approx 0$ \cite{kuk1}$.$ It
is convenient to use, instead of matrix equations, the scalar and vector
pseudo-spin components of the densities and currents as 
\begin{equation}
n=n_{1}+n_{2},{\bf M=}\frac{1}{2}\left(
n_{12}+n_{21},in_{21}-in_{12},n_{1}-n_{2}\right)   \label{eq14}
\end{equation}
and the same for the currents $J_{i}$ and ${\bf J}_{i}$, respectively. In
these pseudo-spin notations, the ''magnetic field'' or ''Larmor frequency''
has the form 
\begin{equation}
{\bf \Omega }_{L}=\left( 0,0,\delta U\right) .  \label{eq15}
\end{equation}
In $^{87}Rb$ this field is low, $\Omega _{L}/T\ll 1$. Here, in contrast to
experiments with true spins, it is almost impossible to change the {\em %
direction }of the magnetic field without exciting the Rabi transitions by 
{\em rf} pulses.

In the hydrodynamic limit of small gradients, equations of motion (\ref{eq10}%
) -(\ref{eq12}) reduce to the well-known Leggett equation for the magnetic
moment ${\bf M}\left( t,{\bf r}\right) $ of the spin-polarized gas (see, 
{\it e.g.,} \cite{mey1,mey2} and references therein): 
\begin{equation}
\partial _{t}{\bf M}+{\bf \Omega }_{L}\times {\bf M}+\partial _{k}{\bf J}%
_{k}=0,  \label{eq17}
\end{equation}
\begin{eqnarray}
\partial _{t}{\bf J}_{k}+{\bf \Omega }_{L}\times {\bf J}_{k}+2g{\bf M}\times 
{\bf J}_{k}+\left( T/m\right) {\bf j}_{k} &=&-{\bf J}_{k}/\tau ,
\label{eq19} \\
{\bf j}_{k} &=&\nabla _{k}{\bf M+M}\nabla _{k}U/T+\left( n/4T\right) \nabla
_{k}{\bf \Omega }_{L}  \label{eq21}
\end{eqnarray}
where $U=\left( 1/2\right) \left( U_{1}+U_{2}+3gn\right) $, $g=4\pi a/m$,
and $\tau $ is the relaxation time (in this case, the difference between
transverse and longitudinal relaxation times is vanishingly small). In
equilibrium, when the currents are absent, ${\bf j}_{k}=0,$ and the
''polarization'' ${\bf M}_{0}$\ follows the field and is directed along it: 
\begin{equation}
{\bf j}_{k}^{\left( 0\right) }\equiv \nabla _{k}{\bf M}_{0}+\left[ {\bf M}%
_{0}\nabla _{k}U+n_{0}\nabla _{k}{\bf \Omega }_{L}/4\right] /T=0.
\label{eq16}
\end{equation}

In Eqs. (\ref{eq17})-(\ref{eq21}), the Larmor frequency can contain weak
coordinate dependence, ${\bf \Omega }_{L}={\bf \Omega }_{0}+\delta {\bf %
\Omega }_{L}\left( {\bf r}\right) $. Under usual assumptions, the terms $%
\partial _{t}{\bf J}_{k}+{\bf \Omega }_{0}\times {\bf J}_{k}\simeq 0$, and
Eq.(\ref{eq19}) yields the following expression for the spin current: 
\begin{eqnarray}
{\bf J}_{k} &=&-\frac{D}{1+\mu ^{2}M_{1}^{2}}\left[ {\bf j}_{k}{\bf +\mu j}%
_{k}\times {\bf M}_{1}+\mu ^{2}{\bf M}_{1}\left( {\bf M}_{1}\cdot {\bf j}%
_{k}\right) \right] ,  \label{eq20} \\
{\bf \mu M}_{1} &=&\tau \left( 2g{\bf M+}\delta {\bf \Omega }_{L}\right) , 
\nonumber
\end{eqnarray}
where $D=\tau T/m$ is the spin diffusion coefficient.

If the magnetic field and trapping potential are uniform, $\delta {\bf %
\Omega }_{L}=0,$ $\nabla _{k}U=0$, Eq. (\ref{eq20}) is exactly the same as
the standard Leggett equation for the spin current.

\section{Formation of domains}

Often, the magnetization in polarized helium can be non-uniform and
noticeably different from its equilibrium field-driven value. The
non-equilibrium polarization can be created by several techniques such as by
flipping the magnetization by $180%
{{}^\circ}%
$ by a proper {\em rf} pulse in a part of a cell, by optical pumping, by
connecting two cells with different magnetizations, {\em etc}. This
initially non-uniform and non-equilibrium polarization can rapidly evolve,
by means of non-dissipative coherent spin currents, into a quasi-equilibrium
state, which will then slowly relax towards the real equilibrium. Such a
quasi-equilibrium state has been thoroughly studied theoretically and
experimentally by Fomin and Dmitriev \cite{fodm}. They demonstrated that a
large initial non-equilibrium polarization ${\bf M}\left( {\bf r}\right) $
can lead to the coherent formation of longitudinal magnetization domains
with spatially separated up and down spins. The effect was observed in low
temperature $^{3}He-\,^{4}He$ mixtures and normal liquid $^{3}He$. The
existence of such domains requires that the width of the domain wall $%
\lambda $, 
\begin{equation}
\lambda ^{3}\sim \frac{D}{\mu M\left| \nabla \Omega _{L}\right| },
\label{f1}
\end{equation}
should be smaller than the system size, $\lambda \ll L$. Diffusive
dissipative processes lead eventually to the homogenization of the system
and the disappearance of domains. Numerical analysis the domain formation
can be found also in Ref. \cite{Rag1}. Note, that the field gradient $\nabla
\Omega _{L}$ is crucial for maintaining the quasi-stationary domain state;
without this gradient, a domain-like distribution of magnetization becomes
unstable with respect to transverse perturbations (see the next section).

Though the main equations for spin-polarized helium and two-component alkali
gases are similar, there are some differences in domain formation between
helium and JILA experiments. In JILA\ experiment \cite{jila1}, the initial
condition corresponds to the large transverse magnetization $M_{tr}\sim n/2$
created by a $90%
{{}^\circ}%
$ {\em rf} pulse. In experiments and calculations of Refs. \cite{fodm}, the
initial state is longitudinal. However, since the formation of domains in
Refs. \cite{fodm} goes through large transverse currents, the
quasi-equilibrium domain state is the same. A more important difference is
that the experiments on alkali gases are performed only in small-size traps.
Taking into account the trap profile $U\left( x\right) \simeq m\omega
_{0}^{2}x^{2}/2$ and that $\mu =2g\tau $, $D=T\tau /m,$ $\delta \Omega
_{L}=m\nu ^{2}x^{2}/2$, we get that in the trap of size $x\sim L=\sqrt{%
T/m\omega _{0}^{2}}$, the width of the domain wall is 
\begin{equation}
\frac{\lambda }{L}\simeq \left[ \frac{1}{4\pi anL^{2}}\frac{\omega _{0}^{2}}{%
\nu ^{2}}\right] ^{1/3}.  \label{f3}
\end{equation}
The typical values of the parameters, $\tau \approx 10^{-2}$ $s$, $n\approx
4\cdot 10^{13}$ $cm^{-3}$, $a=5\cdot 10^{-7}$ $cm$, $m=1.5\cdot 10^{-22}$ $g$%
, $\omega _{0}\sim 7\ Hz,$ combined with an estimate $\nu \lesssim 0.1\ Hz$,
lead to 
\begin{equation}
\lambda /L\gtrsim 0.34.  \label{f4}
\end{equation}
Therefore, it is conceivable that the separation of components in the JILA
experiment is similar to the formation of three Fomin-Dmitriev domains. This
picture of domain formation is consistent with the results of Refs. \cite
{lev1,lal1,wil1}. The temporal symmetry of formation and disappearance of
the domains is explained by the fact that in JILA\ experiments $\mu M$ is
not very large and the coherent processes, which are responsible for the
formation of the domains, have approximately the same time constants as
diffusion, which is responsible for the homogenization. Note, that because
of the peculiar initial condition - large transverse polarization - the
formation of domains does not require the instability in transverse channel
as it is often necessary when starting from a non-equilibrium longitudinal
configuration (see below).

There is another, earlier unexplained experimental fact. The increase of the
field gradient $\nu $ resulted in threshold stratification at some value $%
\nu >0.18\ Hz$ \cite{jila1}. The above estimates show that at $\nu \simeq
0.2\ Hz$ the number of domains can jump to five and give the impression of
such stratification. Since the diffusion equilibration accelerates rapidly
with an increase in the number of domains, this can also produce an
impression of instability in the experiment.

\section{Transverse instability: Application to the JILA experiment}

It is worth noting that the dynamics of the domain formation in the
experiment can often start as {\it transverse instability} in accordance
with Castaing's mechanism \cite{cas1}. Let us, first, briefly describe the
Castaing instability in spin-polarized helium systems and then look at the
implications for the JILA-type experiments.

Suppose the system is homogeneously polarized with uniform magnetization $%
{\bf M}_{0}$. If one is interested in small fluctuations $\delta {\bf M}$ in
transverse channel that are perpendicular to the equilibrium magnetization $%
{\bf M}_{0}$, the linearized Eqs. (\ref{eq17}), (\ref{eq20}) describe the
spectrum of the circularly polarized Silin spin waves, 
\begin{equation}
\omega =\Omega _{0}+\frac{Dk^{2}}{1+\mu ^{2}M_{0}^{2}}\left( i-\mu
M_{0}\right) .  \label{si1}
\end{equation}

When the magnetic moment is driven from the equilibrium and acquires a
gradient ${\bf \nabla }M_{0}$ without change in direction, it is often
assumed that this inhomogeneity will slowly disappear as a result of some
relaxation process. This is not always the case. Often, the magnetic moment
develops spontaneously a noticeable transverse component, which, in turn,
may cause large longitudinal spin currents. Final relaxation towards the
equilibrium is much slower. Castaing \cite{cas1} showed that a purely
longitudinal state with a non-equilibrium gradient of magnetization becomes
unstable with respect to transverse fluctuations $\delta {\bf M}$. The
instability occurs with respect to longwave transverse fluctuations with the
wave vectors $k^{2}<2\mu {\bf k\cdot \nabla }M_{0}$. Mathematically this
means that the imaginary component of the spectrum (\ref{si1}) acquires,
according to Eq. (\ref{eq20}), the additional term,

\begin{equation}
{\rm Im\ }\omega ={\frac{D}{1+\mu ^{2}M_{0}{}^{2}}}(k^{2}-2\mu {\bf k\cdot
\nabla }M_{0}{\frac{\mu ^{2}M_{0}{}^{2}}{1+\mu ^{2}M_{0}{}^{2}}}).
\label{eq400}
\end{equation}
The imaginary part of the spectrum of transverse fluctuations near the
distribution ${\bf M}_{0}$ becomes negative when

\begin{equation}
k^{2}<2\mu {\bf k\cdot \nabla }M_{0}{\frac{\mu ^{2}M_{0}{}^{2}}{1+\mu
^{2}M_{0}{}^{2}}}  \label{eq220}
\end{equation}
The change in sign of ${\rm Im\ }\omega $\ signals the instability. Since
the direction of the vector ${\bf k}$ is not fixed, this condition can
always be satisfied in an infinite system for sufficiently longwave
fluctuations. In polarized quantum gases, with the exception of small
polarizations, the spin rotation parameter $\mu M$ is not small, $\mu M>1$.
Then the condition $\left( \ref{eq220}\right) $ is roughly equivalent to 
\begin{equation}
k^{2}<2\mu {\bf k\cdot \nabla }M_{0}.  \label{eq221}
\end{equation}

This type of instability in transverse channel was observed in several
experiments with $^{3}He\uparrow $, Refs.\cite{aki,nun,owe}. Though\ the
conditions and the setups varied, the experiments were performed for the
systems in which the coherent effects exceeded the diffusion damping, $\mu
M_{0}>1.$ Usually, the initial condition corresponded to the large gradient
of longitudinal magnetization that could be created, for example, by a $180%
{{}^\circ}%
$ pulse in a part of the cell. The instability started in accordance with
Castaing criterion and often led to long-lived NMR ringing. Though the
Castaing instability in transverse channel often serves as an initial stage
of the domain formation, the instability, by itself, is not necessary for
the domain formation\cite{Rag1}.

The experiment with trapped $^{87}Rb$ \cite{jila1} was performed in such a
way so that the Castaing instability was not displayed. First, the
polarization in the initial state was already transverse thus eliminating
the need for strong transverse currents on the initial stage of the domain
formation. Second, this transverse magnetization, probably, did not have
large non-equilibrium gradients. However, the Castaing instability is within
an easy reach and can be observed in a slightly modified experiment.

In JILA-type experiments, the analog of the spin rotation parameter $\mu
M_{0}$ is $\tau gM_{0}$. Numerically, this parameter is $\mu M_{0}=\tau
gM_{0}\sim 17\Delta $, where $\Delta =M_{0}/n$ is the degree of pseudo-spin
polarization.$\ $For trapped alkali gases, it is difficult to achieve a high
degree of the pseudo-spin polarization using the so-called brute force
technique, {\it i.e.,} the effective ''magnetic field'' $\Omega _{L}$. On
the other hand, the preparation of the gas with a high non-equilibrium
polarization is even simpler than in helium and does not require elaborate
techniques such as optical pumping, Castaing-Nozieres method, {\it etc. }For
example, one can simply start from a pure component.

The instability should be re-examined for trapped alkali gases for which the
confining potential $U$ imposes an additional limitation. The reason is that
in a system of the size $L$, the wave vectors smaller than $1/L$ are absent
and the condition (\ref{eq220}) must be considered in conjunction with the
requirement $k>1/L$. On the other hand, the polarization gradient $\nabla
M_{0}$ introduces another length scale, $L_{M}\sim M_{0}/\nabla M_{0}$,
which, depending on the magnitude of the gradient and the length over which
it is spread, can be larger or smaller than $L$. As a result, the
instability condition in finite traps acquires the following threshold with
respect to the trap size:

\begin{equation}
L_{M}/L<{\frac{2\mu ^{3}M_{0}{}^{3}}{1+\mu ^{2}M_{0}{}^{2}}}.  \label{eq2500}
\end{equation}
Roughly, the range of the ''dangerous'' wave vectors is $1/L<k<2\mu
M_{0}/L_{M}$.

For the instability to develop, the diffusion equilibration of the gradients
should not be much faster than the time $t_{C}$ on which the instability
develops. The diffusion relaxation of a longitudinal gradient $\nabla {\bf M}%
_{0}$ takes place on a time scale of the longitudinal spin diffusion. The
lower limit for this time can be estimated as $t_{D}\sim L^{2}/D$. In
certain situations of open systems the non-equilibrium gradients can be
supported even on much longer times. Assuming that $\mu M_{0}$ is large and $%
k$ is far above the threshold, the estimate of the time $t_{C}\sim 1/|{\rm %
Im\,}\omega |$ from Eq.(\ref{eq400}) yields

\begin{equation}
t_{C}\sim L_{M}^{2}/D.  \label{eq401}
\end{equation}
Comparison with the diffusion time shows that the instability in small traps
can be observed only if 
\begin{equation}
L>L_{M}.  \label{eq402}
\end{equation}
This is the simplest condition for the observation of the instability in
finite isolated traps.

If $L\sim L_{M},$\ the diffusion time is comparable to the time of
development of instability. However, since at $\mu M\gg 1$ the instability
is developed at relatively large wave vectors $k>1/L$, the signs of the
instability should be still clearly visible in experiment as higher
harmonics in magnetization distribution in transverse channel, $i.e$.,
''ringing'' similar to that observed in helium experiments, Refs.\cite
{aki,nun,owe}.

Above we neglected the possibility of having gradients of magnetic field
which can stabilize the non-uniform distribution of magnetization \cite
{fodm,Rag1}. In order to observe the instability in the presence of the
field gradient, Eq.(\ref{eq402}) for $L_{M}$ should be supplemented by the
condition that $L_{M}$ is also much smaller than the width of the
quasi-stationary domain wall $\lambda $, Eq. (\ref{f1}), 
\begin{equation}
L_{M}\ll \left( \frac{D}{\mu M\left| \nabla \Omega _{L}\right| }\right)
^{1/3}\sim L\left[ \frac{1}{4\pi anL^{2}}\frac{\omega _{0}^{2}}{\nu ^{2}}%
\right] ^{1/3}.  \label{eq403}
\end{equation}
According to the estimates from the previous section, $\lambda \gtrsim L/3$
for the field gradients $\nu \lesssim 0.1\ Hz$. This means that if the field
gradients are noticeably smaller than this value, Eq. (\ref{eq403}) does not
impose any new restrictions for the observation of the instability.

Note, that in the situations, when the non-equilibrium gradient is supported
externally, {\it i.e.}, when the quasi-steady currents can flow, the
condition (\ref{eq402}) is not required. Under such conditions, the scale $%
L_{M}$ is $L_{M}\sim L$ and the instability develops, as in helium systems,
as long as $\mu M_{0}\gg 1$.

There are many experimental ways for creating a non-equilibrium gradient of
polarization of the trapped gases sufficient for the observation the
Castaing instability, Eq. (\ref{eq402}). As a result of this instability,
the polarized system should acquire a rapidly growing perpendicular
component, ${\bf M}_{tr}\bot {\bf M}_{0}$. The dynamics of similar processes
in helium, though in very different geometries, has been studied
computationally in Ref. \cite{Rag1}.

In experiment \cite{jila1}, the $90%
{{}^\circ}%
$ {\it rf}-pulse created the almost {\it uniform} {\it transverse}
polarization ${\bf M}_{0}$ in, {\it i.e.}, the $x$-direction in the
pseudo-spin space. The spin-rotation parameter was relatively large,

\begin{equation}
\mu M_{0}\approx \tau gn/2\sim 9.  \label{eq300}
\end{equation}
[The trapping potential was responsible for some non-uniformity of
polarization. This {\it equilibrium} non-uniformity could not lead to any
instability; actually, small field gradients help to stabilize the system].
The small fluctuations $\delta {\bf M}$ around this configuration were, if
one ignores the small magnetic field, the gapless Silin waves, Eq. (\ref{si1}%
), and there was no instability. However, if the initial polarization $M_{x}$
in experiment \cite{jila1} were sufficiently non-uniform for the instability
to occur, the observed component separation , {\it i.e.,} the formation of
domains with different $M_{z}$ would have taken place much faster than in
Refs. \cite{jila1,lev1,lal1,wil1}. This can be checked experimentally.

There are several other possible experiments with the instability and domain
formation that can be performed with trapped alkali gases along the lines of
the helium experiments. First, one can start from a one-component gas and
flip the ''magnetization'' by $180%
{{}^\circ}%
$, {\em i.e.,} excite the Rabi transition, in a part of a trap. Then one can
observe long time magnetization ringing in the system similar to Refs.\cite
{aki,nun,owe}. In this experiment, however, the transverse polarization
might already exist after the pulse on the interface between the components
due to some residual non-uniformity of the Rabi frequency \cite{jila2}. In
this sense, it is not clear whether the instability can be unambiguously
identified.

The next experiment involves bringing together two clouds consisting of the
two different components, which have no memory of each other. Employing the
pseudo-spin language, this two clouds are characterized by the opposite
polarizations in the $z$-direction with no transverse components. Then, the
dynamics of mixing of this two clouds can proceed along the line of Castaing
instability, as opposed to a trivial (incoherent) diffusion mixing.
Accordingly, the transverse non-uniform magnetization will arise
spontaneously near the interface between the mixing components. Then it will
slowly relax till the equilibrium state is reached. In some sense, this
effect is the time reversal analog of the experiment \cite{jila1}. Indeed,
in the case \cite{jila1}, the initial non-equilibrium transverse
polarization resulted in the transient separation of the components. The
effect of Castaing instability results in the {\it spontaneous} creation of
the transverse polarization in the process of the components mixing.

Another possibility is to create a non-uniform stable distribution of
longitudinal polarization in which the criterion (\ref{eq402}) is violated
and then to change rapidly the size of the trap so that to comply with the
instability condition (\ref{eq402}).

The last suggestion is more speculative. The non-uniformity of the initial
non-equilibrium configuration can be caused by the sensitivity of the
two-photon Rabi transition to the gradients of difference of the trapping
potentials \cite{jila2}, that is, to the value of the Larmor frequency $%
\Omega _{L}$. This spacial non-uniformity of the pulse frequency $\Omega
\left( {\bf r}\right) $ results in formation of helicoidal structures after
the pulse. For example, if the Rabi rotation takes place around the $x$%
-axis, the structure of magnetization ${\bf M}\left( {\bf r}\right) $ after
the pulse of duration $t$ has the form $M_{z}+iM_{y}=M_{0}\exp (i\Omega (%
{\bf r})t).$ Such helicoidal polarization ${\bf M}\left( {\bf r}\right) $
has large gradients and may be unstable. As was emphasized in Ref.\cite
{jila2}, the Rabi frequency gradient $\nabla \Omega $ can be controlled with
high accuracy. Accordingly, one can control the magnetization gradients.
However, the time evolution of the helicoid-like structure ${\bf M}\left( 
{\bf r}\right) $ is described by a convoluted set of coupled transverse and
longitudinal equations. At present, it is not clear to what extent the
instability condition for such helicoids differs from the above equations
for simple geometries. This general suggestion can be modified in various
ways so that to create different structures some of which can be even one
dimensional and are easier to describe theoretically than the helicoids of a
general form. However, a practical implementation strongly depends on an
available experimental setup.

In most of the suggested experiments, the detection of the instability and
the spontaneous formation of the transverse magnetization, {\em i.e.}, of
the large off-diagonal components of the density matrix associated with
continuous Rabi transitions, can be done using the Ramsey spectroscopy \cite
{ram1}.

\section{Conclusions}

In conclusion, we pursued the analogy between the two-component $^{87}Rb$
gases and spin-polarized helium systems. The observed separation of
components in experiment \cite{jila1} is analogous to the formation of
coherent precessing domains in helium \cite{fodm}. This interpretation is
consistent with the conclusions of Refs.\cite{lev1,lal1,wil1}. The width of
the domain wall agrees with the formation of up to three domains. An
increase in the gradient of the pseudo-magnetic field can result in a
spontaneous stratification of the system. A polarization gradient in the
initial state can drastically speed up the domain formation. Apart from the
observation of the domain formation, there are several other interesting
experimental possibilities associated with the pseudo-spin Castaing
instability in transverse channel. This instability is almost inevitable in
highly polarized systems with non-uniform distribution of the pseudo-spin
and can be observed in the JILA-type experimental setups. The conditions for
this instability in finite traps are delineated.

The authors acknowledge support by the CUNY grant PSC-63499-0032 (A.K.) and
by the NSF grant DMR-0077266 (A.M.).

\end{document}